\documentclass[aps,prx,superscriptaddress,twocolumn,amsmath,longbibliography]{revtex4-2}

\usepackage[utf8]{inputenc}
\usepackage{longtable}
\usepackage{amsmath}%
\usepackage{amsfonts}%
\usepackage{amssymb}%
\usepackage{graphicx}
\usepackage[colorlinks=true,linkcolor=blue,urlcolor=black,hyperfootnotes=true,citecolor=blue]{hyperref}
\usepackage{ulem} 
\usepackage[product-units=power, separate-uncertainty=true, multi-part-units=single]{siunitx} 
\usepackage[usenames,dvipsnames]{xcolor}
\usepackage{threeparttable}
\usepackage{multirow}
\usepackage{kantlipsum,setspace}

\newcommand{\dIdU}{$\dd I / \dd V$}
\newcommand{\dd}{\text{d}}
  
\newcommand{\BiSe}{${\mathrm{Bi}}_{2}{\mathrm{Se}}_{3}$}

\newcommand{\PSBS}{$(\mathrm{PbSe})_{5}(\mathrm{Bi}_{2}\mathrm{Se}_{3})_{6}$}

\def\app#1#2{%
  \mathrel{%
    \setbox0=\hbox{$#1\sim$}%
    \setbox2=\hbox{%
      \rlap{\hbox{$#1\propto$}}%
      \lower1.1\ht0\box0%
    }%
    \raise0.25\ht2\box2%
  }%
}

\newcommand{\etal}{\textit{et al.}}

\usepackage{array}
\newcommand{\PreserveBackslash}[1]{\let\temp=\\#1\let\\=\temp}
\newcolumntype{C}[1]{>{\PreserveBackslash\centering}p{#1}}
\newcolumntype{R}[1]{>{\PreserveBackslash\raggedleft}p{#1}}
\newcolumntype{L}[1]{>{\PreserveBackslash\raggedright}p{#1}}

\begin{document}

\title{Quasiparticle interference on the surface of Bi$_{\mathbf{2}}$Se$_{\mathbf{3}}$ terminated (PbSe)$_{\mathbf 5}$(Bi$_{\mathbf 2}$Se$_{\mathbf 3}$)$_{\mathbf 6}$}

\author{Mahasweta Bagchi} \affiliation{Physics Institute II, University of Cologne, D-50937 K{\"o}ln, Germany}

\author{Philipp R{\"u}{\ss}mann}
\affiliation{Peter Gr{\"u}nberg Institut  (PGI-1), Forschungszentrum J{\"u}lich and JARA, 52425 J{\"u}lich, Germany}
\affiliation{Institute for Theoretical Physics and Astrophysics, University of W{\"u}rzburg, 97074 W{\"u}rzburg, Germany}

\author{Gustav Bihlmayer} \email{g.bihlmayer@fz-juelich.de} \affiliation{Peter Gr{\"u}nberg Institut  (PGI-1), Forschungszentrum J{\"u}lich and JARA, 52425 J{\"u}lich, Germany}

\author{Stefan Bl{\"u}gel} \affiliation{Peter Gr{\"u}nberg Institut  (PGI-1), Forschungszentrum J{\"u}lich and JARA, 52425 J{\"u}lich, Germany}

\author{Yoichi Ando }   \affiliation{Physics Institute II, University of Cologne, D-50937 K{\"o}ln, Germany}

\author{Jens Brede} \email{brede@ph2.uni-koeln.de} \affiliation{Physics Institute II, University of Cologne, D-50937 K{\"o}ln, Germany}

\date{\today}

\begin{abstract}
 Among the family of topological superconductors derived from {\BiSe}, $\mathrm{Cu}_x(\mathrm{PbSe})_{5}(\mathrm{Bi}_{2}\mathrm{Se}_{3})_{6}$ is unique in its surface termination of a single quintuple layer (QL) of the topological insulator (TI) \BiSe{} on an ordinary insulator PbSe.
Here, we report a combined scanning tunneling microscopy (STM) and density functional theory (DFT) characterization of the cleaved surface of the parent compound $(\mathrm{PbSe})_{5}(\mathrm{Bi}_{2}\mathrm{Se}_{3})_{6}$ (PSBS). Interestingly, the potential disorder due to the random distribution of native defects is only $\Gamma \sim \qty{4}{meV}$, comparable to the smallest reported for TIs. Performing high-resolution quasiparticle interference imaging (QPI) near the Fermi energy ($E-E_\mathrm{F} = \qtyrange{-1}{0.6}{\eV}$) we reconstruct the dispersion relation of the dominant spectral feature and our \textit{ab initio} calculations show that this surface feature originates from two bands with Rashba-like splitting due to strong spin-orbit coupling and inversion symmetry breaking. Moreover, only a small hexagonal distortion of the calculated Fermi surface is seen in the full momentum space distribution of the measured scattering data. Nevertheless, the scattering pattern at lower energies transforms into a flower-like shape with suppressed intensity along the $\overline{\Gamma \mathrm{K}}$ direction. We show that this effect is not due to the forbidden backscattering in the spin-momentum locked surface state in Bi$_2$Se$_3$ but reflects the threefold symmetry of the scattering potential.
\end{abstract}

\maketitle
\newpage

\section{Introduction}
The family of bulk superconductors derived from doped \BiSe{} show overwhelming experimental evidence for topological superconductivity (SC) in the bulk~\cite{Sato2017} but even the observability of signs of any SC at the surface remains highly controversial~\cite{Bagchi2022}. In particular, the absence of conclusive experimental evidence of SC in the topological surface state (TSS) in the same measurement~\cite{Wilfert2018,Bagchi2022}, calls for further experimental and theoretical work.
To this end, superconducting Cu-doped \PSBS{} (PSBS) is particularly interesting, first, because it clearly shows signatures of odd-parity spin-triplet-like superconductivity in the bulk~\cite{Andersen2018,Andersen2020} \textit{and} at the surface~\cite{bagchi2023}.
Second, it offers a platform to tune and control the properties of the TSS; surface terminations consisting of a single \BiSe{} quintuple layer (QL) host a quasi-parabolic band, whereas two or more QLs show a hybridized and Rashba-split TSS with a sizeable gap at the Dirac point~\cite{Nakayama2012,Nakayama2019}, but no spin-resolved data have been obtained.
First-principles calculations based on density functional theory (DFT) by Momida \textit{et al.}~\cite{Momida2018} elucidated that in the bulk of PSBS, due to a high density of Bi antisite ($\mathrm{Bi}_\mathrm{Pb}$) defects in the PbSe layer, bands stemming from PbSe are pushed down from the Fermi energy, leaving the topological interface states of {\BiSe} having a gapped Dirac-cone-like dispersion at the Fermi level. However, neither surface nor spin properties were addressed in these DFT calculations.

It has been elucidated that in the topological surface state, hexagonal warping~\cite{Fu2009} enables scattering along the $\overline{\Gamma \mathrm{M}}$ direction, leading to flower-like patterns in the Fourier transforms (FTs) of the real-space QPI data~\cite{Beidenkopf2011}.
Here we performed QPI imaging at the surface of PSBS and observed a transition from a hexagonal pattern at the Fermi level to a flower-like shape at \textit{lower} energies. Our \textit{ab initio} calculations show that the bandstructure at these energies is quasi-parabolic with sizeable Rashba-like splitting and only a small hexagonal distortion at the Fermi level. Therefore, the observed flower-like pattern is not due to the band structure but due to the Bragg condition imposed by the typical triangular geometry of the scattering sites. We recover good agreement between theory and experiment by modeling the experimental data as a convolution of an analytic approximation of the threefold symmetric scatterers and the calculated band structure.

\section{Methods}

\subsection{Experimental Methods}
\textit{Crystal growth.} We grew $(\mathrm{PbSe})_{5}(\mathrm{Bi}_{2}\mathrm{Se}_{3})_{6}$ single crystals using a modified Bridgeman method as described previously \cite{Sasaki2014,Andersen2018}.

\textit{STM measurements.} STM experiments were carried out under UHV conditions with a commercial system (Unisoku USM1300). Data were acquired at $1.7$~K unless mentioned otherwise. Topograph and \dIdU{} maps were recorded in the constant-current mode. Point spectroscopy data was obtained by first stabilizing for a given set-point condition and then disabling the feedback loop. \dIdU{} curves were then recorded using a lock-in amplifier by adding a small modulation voltage $V_{\text{mod}}$ to the sample bias voltage $V$. 
We have used both PtIr and W probe tips. All PtIr tips used were commercially obtained from Unisoku. The W tips were made in-house. Both types were electrochemically etched. The PtIr tips were either fresh new tips or they were prepared by standard Ar ion etching (at an argon pressure of $3\times10^{-6}$~mbar and a voltage of $1$~kV), followed by repeated heating by electron bombardment ($\sim15$~W) for $20$~s. Further tip forming was done by scanning on the Cu(111) surface until a clean signature of the surface state was obtained in spectroscopy. The PSBS crystal was cleaved by knocking off a 10~mm sized pole glued on the sample. The two-component epoxy glue (EPO-TEK H21D) was hardened by heating to $373$~K. STM data were processed using Igor Pro.

\subsection{DFT calculations}

To investigate the electronic structure, we performed density functional theory (DFT) calculations in the local density approximation~\cite{Vosko:80} using the full-potential linearized augmented plane-wave method as implemented in the {\sc Fleur} code~\cite{Fleur}. The symmetric film consisted of two vertical units, which included four quintuple layers of Bi$_2$Se$_3$ and two PbSe layers embedded in semi-infinite vacuum~\cite{Krakauer:79}, leading to a total of 160 atoms per unit cell (see Fig.~\ref{fig:fig1} a-c). We substituted Bi atoms for Pb, as in Ref.~\cite{Momida2018} to simulate the Pb off-stoichiometry. We used the virtual crystal approximation to avoid artifacts from the quasi-regular arrangement of these Bi$_{\textrm{Pb}}$ atoms. The structure was relaxed using a $1 \times 5$ regular $k$-point grid and the product of the smallest muffin-tin radius, $R_{\rm MT}$, with the basis set cutoff, $K_{\rm max}$, was $8.6$. For the band structures, DOS calculations and Fermi surfaces, spin-orbit coupling was included self-consistently. In the DOS calculations, a $6 \times 15$ $k$-point grid was used. For the band structure, the Brillouin zone was unfolded to the simple Bi$_2$Se$_3$ unit cell and the Fermi surface was sampled with 6080 $k$-points in the corresponding irreducible zone.
 
\section{Results}
\label{sec:results}
\begin{figure}[h!]
    \includegraphics[width=0.90\columnwidth]{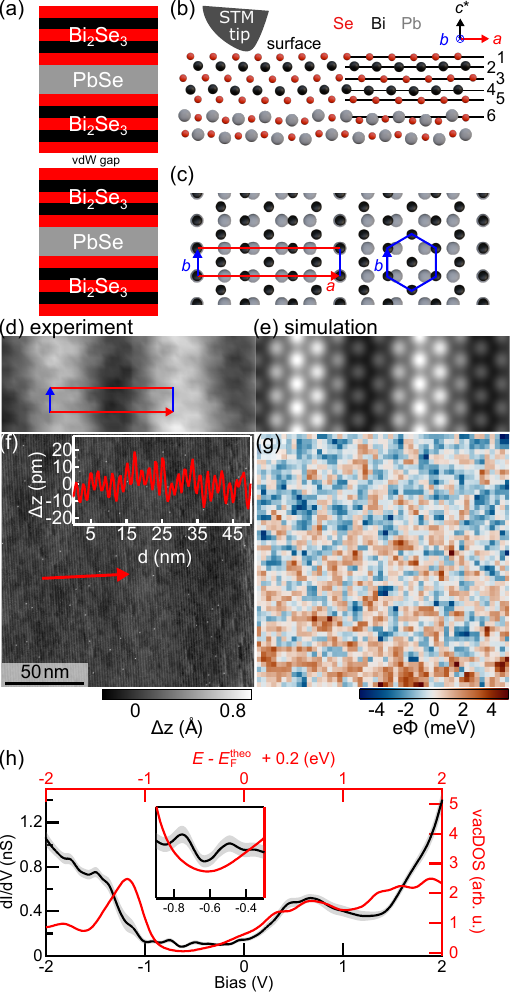}
	\caption{Schematic representation (a) of PSBS consisting of Bi$_2$Se$_3$ and PbSe units. Side view (b) of the surface atomic layers; the top view (c) of only the hexagonal Bi$^{(2)}$ and square Pb$^{(6)}$ lattices illustrating the moir{\'e} effect underpinning the quasi-one-dimensional stripe-pattern. Superscripts indicate the layer index. Atomically resolved STM topography  [(d), $I_0=\qty{20}{\nA}$, $V_0=\qty{900}{\mV}$] and simulated local DOS (e) for energies [$E_{\mathrm F}, E_{\mathrm F}+0.9$ eV] at a distance of \qty{6.35}{\AA} above the surface. (f) Large-scale topography ($I_0=\qty{1}{\nA}, V_0=\qty{2}{\V}$) with a line cut along the red arrow shown as an overlay. Scale bar is \qty{50}{\nm}. (g) The spatial distribution of the electrostatic potential $e \Phi({\bf r}) $ of the area shown in (f) was derived from a $50 \times 50$ point STS grid ($I_0=\qty{1}{\nA}, V_0=\qty{2}{\V}$). (h) Comparison of the average spectrum of the STS grid (black trace, the grey area denotes the standard deviation) and the vacuum DOS (red trace) determined from first principles calculations at \qty{12.6}{\AA} above the surface.}
	\label{fig:fig1}
\end{figure}

PSBS consists of blocks of two quintuple layers (QLs) of the prototypical topological insulator Bi$_2$Se$_3$ and a single bilayer of the trivial insulator PbSe as schematically depicted in Fig.~\ref{fig:fig1}~(a).
 The PSBS crystal cleaves in the van der Waals gap between the Bi$_2$Se$_3$ QLs, exposing large flat terraces of a single QL [Fig.~\ref{fig:fig1}~(f)]. Interestingly, a prominent one-dimensional stripe pattern is visible in the overview image, and from the line cut we quantify the period to about $50/24 \approx \qty{2.1}{nm}$. The structural origin of this quasi-one-dimensional pattern becomes clear when performing atomic resolution imaging [Fig.~\ref{fig:fig1}~(d)].
A bright stripe running along the crystallographic $b$-axis repeats roughly every \qty{2.1}{nm}, i.e., every 6 Se-ion rows.
By comparison with the crystallographic unit cell [Fig.~\ref{fig:fig1}~(b) and (c)] the observed contrast is understood as a moir{\'e} effect arising from the hexagonal \BiSe{} lattice on top of the square PbSe one. The stripe pattern is also found in the simulated STM image [Fig.~\ref{fig:fig1}~(e)] derived from our first-principles calculation indicating good agreement with the experiment.

Next, we have taken a grid of $50 \times 50$ \dIdU -spectra covering the area shown in Fig.~\ref{fig:fig1}~(f). The numerical average of all spectra [black trace in Fig.~\ref{fig:fig1}~(h)] shows a steep increase around \qty{-1}{V} and \qty{0.2}{V} as well as a minimum at around \qty{-0.65}{V}.
We note a good overall agreement with the vacuum LDOS extracted from our first-principles calculations (red trace) when accounting for a rigid shift of the Fermi energy $\left( E_{\mathrm{F}}^{\mathrm{theo}}+\qty{0.2}{eV}=E_{\mathrm{F}}^{\mathrm{exp}} \right)$.
If not stated otherwise, from now on we use $E_\mathrm{F}^{\mathrm{exp}}$ as the Fermi energy throughout the manuscript.
Interestingly, the spectral features below \qty{-1}{V} and above \qty{0.6}{V} show significant variations as a function of spatial position, reflected in the large standard deviation (grey area). At the same time, the spectra between \qtyrange{-1}{0.2}{V} are comparatively homogenous and allow the mapping of the onset energy ($eV_\text{min}$) of the main spectral feature by tracing the minimum of the LDOS  to quantify the electrostatic potential disorder~\cite{Skinner2013,Borgwardt2016,Knispel2017} at the surface of PSBS.
We define $e \Phi({\bf r})=e (\langle V_\text{min} \rangle - V_\text{min}({\bf r}))$, with the average energy and the energy at position ${\bf r}$ given by $e\langle V_{\mathrm{min}} \rangle$ and $eV_\text{min}({\bf r})$, respectively. In Fig.~\ref{fig:fig1}~(g) we plot the map of the potential disorder $e \Phi({\bf r}) $, showing characteristic excess-electron puddles with negative $e \Phi$ in blue and fewer-electron puddles with positive $e \Phi$ in red. From the statistical analysis~\cite{brede2024} of $e \Phi({\bf r})$ we define the amplitude ($\Gamma \sim \qty{4}{meV}$) of the potential disorder as the width of Gaussian distribution fit to the histogram of $e\Phi$ and the point where the azimuthal average of the 2D-autocorrelation of $e\Phi({\bf r})$ drops more rapidly than $1/r$ as the characteristic length scale of the surface puddles ($l \sim \qty{5}{\nm}$). The obtained $\Gamma$ is comparable to the smallest value reported for typical topological insulator single crystals such as Bi$_2$Se$_3$ and Bi$_2$Te$_3$ with $\Gamma=\qtyrange{5}{13}{\meV}$ and $\Gamma\approx\qty{3}{\meV}$, respectively~\cite{brede2024}.

\begin{figure}[h!]
	\includegraphics[width=0.8\columnwidth]{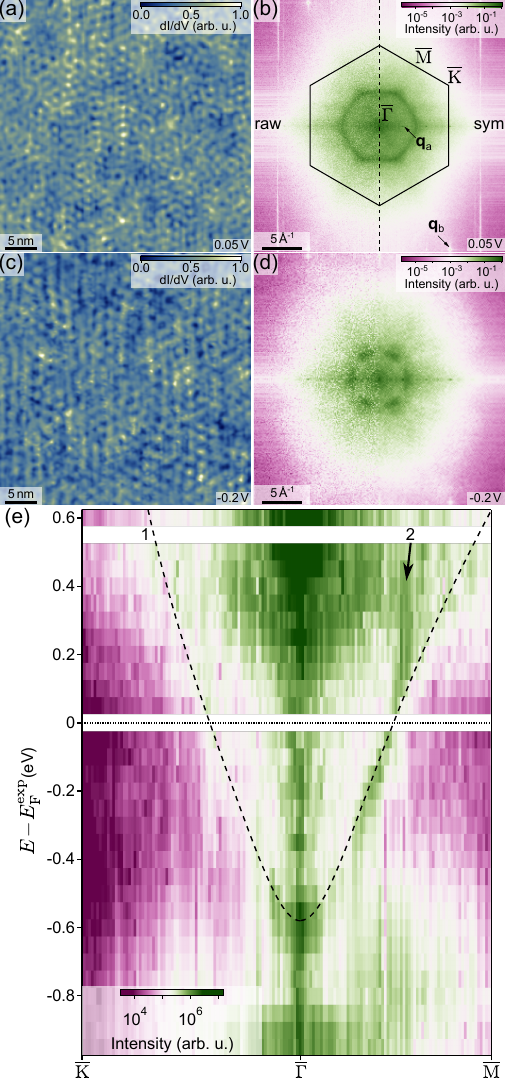}
	\caption{Quasiparticle interference (QPI) in PSBS. Maps of the differential conductance $\dd I/\dd V({\bf r})\approx \mathrm{LDOS}({\bf r})$ for $V_0=\qty{50}{\mV}$ (a) and $\qty{-200}{\mV}$ (c). (a,c) Scale bar \qty{5}{\nm}, $I_0=\qty{2}{\nA}$. (b) The Fourier transform (FT) of (a) before (raw, left half) and after symmetrization (sym, right). The surface Brillouin zone of \BiSe{} is superimposed, and Bragg points due to the quasi-one-dimensional stripe pattern (${\bf q}_a$) and the hexagonal Se-lattice (${\bf q}_b$) are highlighted. (d) The FT of (c) has higher (lower) intensity along the $\overline{\Gamma M}$ ($\overline{ \Gamma K}$) direction. (b,d) Scale bar $\qty{5}{\AA^{-1}}$. (e) The experimental scattering intensity (color map) extracted along the $\overline {K \Gamma M}$ direction and the dispersion relation (black dashed line) given by Eq.~\eqref{eq:TSS} using the scaling relation $q=2k$. For technical details and a deconvolution of the experimental data into features 1 and 2 see Appendix~\ref{app:ExperimentVSBHZmodel}.
	}
	\label{fig:QPImethod}
\end{figure}

After the initial characterization of topographic and electronic features, we characterize the band structure from \qtyrange{-1}{0.6}{V} via quasiparticle interference (QPI) as depicted in Fig.~\ref{fig:QPImethod}.
At first glance, the maps of the differential conductance $\dd I/\dd V({\bf r})$ [Fig.~\ref{fig:QPImethod}~(a,c)] show roughly circular standing wave patterns in the LDOS with increasing wavelength as the bias voltage is reduced. 
The characteristic scattering wave vectors ($q=2\pi/\lambda$) of the periodic features in $\dd I/\dd V({\bf r})$ are conveniently visualized by performing the FT. The resulting $\mathcal{F}_{\dd I/\dd V}({\bf q})$ are plotted in Figs.~\ref{fig:QPImethod}~(b,d).
We improve the signal-to-noise ratio by correcting for thermal drift and symmetrizing our data by mirroring the data along both the $x$ and the $y$ axis. Technical details can be found in Appendix~\ref{app:Driftcor} and the effect of the data treatment is visualized in Fig.~\ref{fig:QPImethod}~(b,d) by plotting both raw (left half) and symmetrized (right half) data.
The $\mathcal{F}_{\dd I/\dd V}({\bf q})$ data show Bragg spots due to the hexagonal lattice (${\bf q}_b$) of the top Se-layer and the rectangular lattice of PSBS (${\bf q}_a$).
As a guide to the eye, we superimposed the hexagonal surface Brillouin zone of Bi$_2$Se$_3$ onto the experimental data. 
A hexagonal feature with a wavevector $q\approx \qty{0.45}{\AA^{-1}}$ is easily identified at \qty{50}{\meV} [Fig.~\ref{fig:QPImethod}~(b)]. The pattern changes into a flower-like shape with $q\approx \qty{0.34}{\AA^{-1}}$ at \qty{-200}{\meV} [Fig.~\ref{fig:QPImethod}~(d)].
In Fig.~\ref{fig:QPImethod}~(e), we plot $\mathcal{F}_{\dd I/\dd V}({\bf q},V)$ along the high-symmetry directions ($\overline{\mathrm{K} \Gamma \mathrm{M}}$) for different tunneling bias voltages. The primary spectral feature exhibits a nearly parabolic dispersion (dashed line). A second feature, which disperses only weakly in momentum as a function of energy, is visible along $\overline{\Gamma M}$; this is spurious and arises from the well-known constant-current set-point effect.\footnote{See Appendix~\ref{app:ExperimentVSBHZmodel} and Mcdonald \etal~\cite{Macdonald_2016} for brief and detailed discussions, respectively, of set-point effects.}

\begin{figure}[h!]
	\includegraphics[width=1.0\columnwidth]{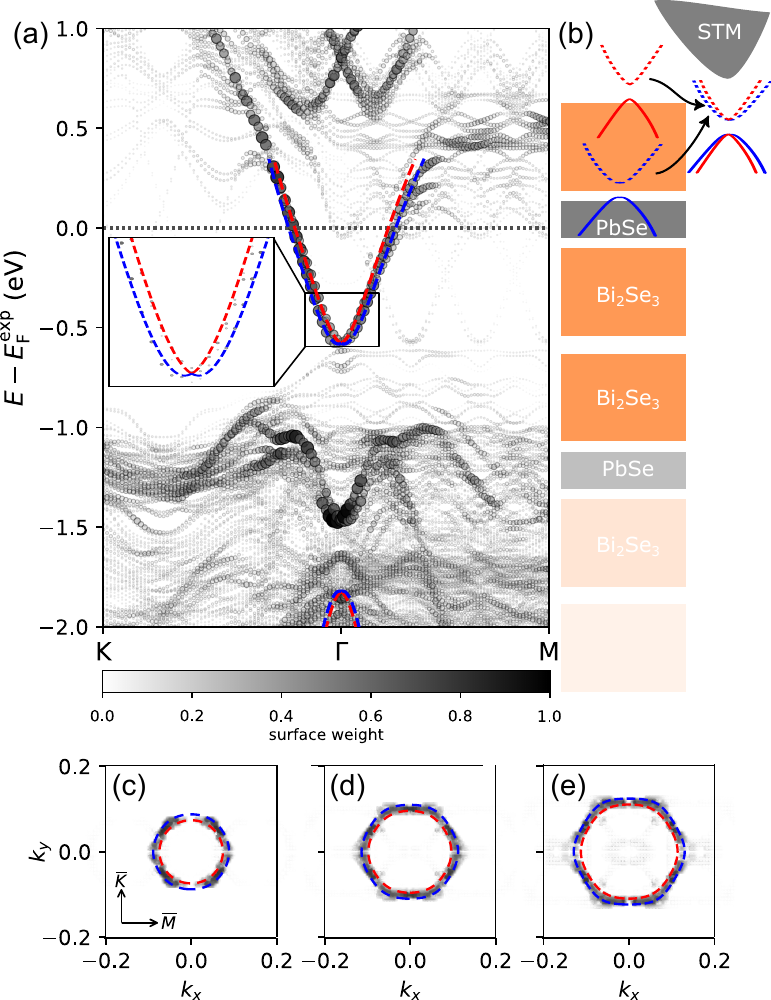}
	\caption{Dispersion relation and constant energy contours from first-principles. (a) Eigenenergies with significant surface weight along the $\overline {K \Gamma M}$ direction from first-principles calculations. (b) Sketch of the interacting surface and interface states from the first Bi$_2$Se$_3$ QL in PSBS where red/blue colors indicate opposite spin. (c,d,e) Constant energy contours (CECs) at $E-E^{\mathrm{exp}}_\mathrm{F}=\qty{-0.2}{\eV}$ (c), \qty{-0.05}{\eV} (d), and \qty{0.05}{\eV} (e). In red (blue for the second spin channel) $E(k)$ and CECs after Eq.~\eqref{eq:TSS} are superposed on the data in (a) and (c-e), respectively.
	}
	\label{fig:DispersionRelationTheory}
\end{figure}

To explain the the main feature in the $\mathcal{F}_{\dd I/\dd V}({\bf q},V)$ data, we turn to the momentum resolved eigenenergies from our first-principles calculations plotted in Fig.~\ref{fig:DispersionRelationTheory}~(a) and recognize the nearly-parabolic feature with high surface weight for $E-E^\mathrm{exp}_\mathrm{F}\approx\qtyrange{-0.6}{0.4}{\eV}$ in good agreement with the experiment. Interestingly, upon close inspection [inset Fig.~\ref{fig:DispersionRelationTheory}~(a) and Appendix~\ref{app:states}] one discerns the contributions of two~\footnote{As discussed in  Appendix~\ref{app:states}, there are contributions of three states, but only two of them have significant surface weight and are relevant for the comparison with the experiment.} states, which have the same energy but are shifted in momentum. The spin-projection of the eigenenergies (Appendix~\ref{app:states}) clearly shows the hallmark of Rashba-SOC, i.e.\ the spin-direction is locked perpendicular to the momentum direction where inner and outer parabolas have opposite helicity. Moreover, and in agreement with previous ARPES measurements~\cite{Nakayama2012,Nakayama2015,Nakayama2019}, we note that the constant energy contours around the Fermi level [Fig.~\ref{fig:DispersionRelationTheory}~(d,e)] have a very small but finite hexagonal distortion.
Based on these observations, we construct a minimal effective 2D continuous model where the top and bottom surface states in thin films of 3D TIs interact~\cite{Shan_2010,Zhang2010} and extend it with the warping term of Ref.~\onlinecite{Fu2009}. This gives the dispersion relation according to~\footnote{Note that with $D=B=0$ for the Hamiltonian of Ref.~\onlinecite{Zhang2010} we arrive at the formulation of the TSS Hamiltonian of Fu~\cite{Fu2009} but without the warping term ($\propto\lambda k^3$).}
\begin{eqnarray}
E_{\sigma}^\pm(k, \theta) &=&E_\mathrm{D} - Dk^2 \pm \bigl[(\tilde{V_z} +\sigma v_\mathrm{D}k)^2 \notag \\
    &&+ \left({\lambda}k^3 {\cos 3\theta}\right)^2 + (\frac{\Delta}{2} - Bk^2)^2\bigr]^{1/2},
    \label{eq:TSS}
\end{eqnarray}
with the Dirac point $E_\mathrm{D}=\qty{-1.2}{\eV}$, Dirac velocity $v_\mathrm{D}=\qty{7.2}{\eV\AA}$, the spin $\sigma=\pm 1$, hexagonal warping $\lambda=\qty{26}{\eV\AA^3}$, hybridization gap $\Delta=\qty{-1.24}{\eV}$, inversion symmetry breaking potential $\tilde{V_z}=\qty{0.05}{eV}$, and the competing quadratic terms $D=\qty{7.8}{eV\AA^2}$, $B=\qty{-8.4}{eV\AA^2}$ that introduce electron-hole asymmetry.
It is important to note that $\tilde{V_z}$ in Eq.~\ref{eq:TSS} introduces a Rashba-like splitting of the top/bottom TSS from the first Bi$_2$Se$_3$ QL as shown in Fig.~\ref{fig:DispersionRelationTheory}(a,b). Interestingly, the symmetry-breaking term at the surface was recently proposed to cause the lifting of the superconducting gap nodes at the surface of CPSBS~\cite{bagchi2023}.
Please note, that we imposed $\Delta B \geq 0$ and $|D|\leq|B|$ in our model to enforce (i) a topological gap \cite{Zsurka2024} and (ii) that the energy gap does not close at larger momenta \cite{Lu2010}, respectively.

Comparison of the minimal model with our calculations shows good agreement for $E-E^\mathrm{exp}_\mathrm{F}\approx\qtyrange{-0.6}{0.4}{\eV}$. We further compare the full momentum distribution of our model with the constant energy contours (CECs) determined from our first-principles calculations for $E-E^\mathrm{exp}_\mathrm{F}=-0.2$, $-0.05$ and $\qty{0.05}{\eV}$ [Fig.~\ref{fig:DispersionRelationTheory}~(c-e)]. Encouragingly, both the nearly circular CEC at \qty{-0.2}{\eV} and the slightly hexagonal CECs near the Fermi level are well captured in the model. Note that the hexagonal distortion of the CECs is a reflection of the small~\footnote{The warping term in Bi$_2$Te$_3$ and Bi$_2$Se$_3$ amounts to $\qty{250}{\eV\AA^3}$ and $\qty{128}{\eV\AA^3}$, respectively~\cite{Fu2009, Kuroda2010}.} warping term in Eq.~\ref{eq:TSS}. We illustrate just how small this contribution is in Fig.~\ref{fig:warping} of the Appendix. 

\begin{figure}[h!]
	\includegraphics[width=0.65\columnwidth]{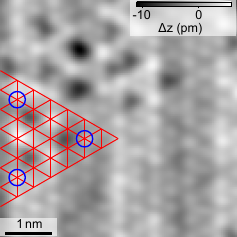}
	\caption{ (a) High-resolution topography of $\mathrm{Pb}_\mathrm{Bi}$-antisite defects ($I_0=\qty{2}{\nA}$, $V_0=\qty{600}{\mV}$). A standard high-pass filter was applied to enhance the contrast. The scale bar is \qty{1}{\nm} and the Se-lattice is partially superimposed in red. Blue circles indicate the position of the three scalar scatters modeled in Eq.~\ref{eq:F_3V}
	}
	\label{fig:DefectSTM}
\end{figure}

\begin{figure}[htb]
	\includegraphics[width=.8\columnwidth]{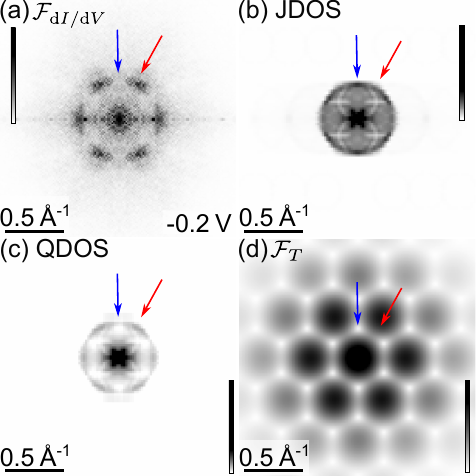}
	\caption{Comparison of calculated JDOS of PSBS with the measured scattering intensity. (a) Symmetrized Fourier transform (FT) of the \dIdU -map taken at \qty{-0.2}{\V} [same data as in Fig.~\ref{fig:QPImethod}~(d)]. The QDOS (c) is the convolution of the JDOS (b) with the FT of the shape function ($\mathcal{F}_{T}$) approximating the triangular scattering potential (d). The scale bar for all panels is \qty{0.5}{\text{\AA}^{-1}} and the color scale of the intensity (arb.~u.) is saturated as indicated to enhance the contrast at the points highlighted by blue and red arrows.}
	\label{fig:qpivsjdos}
\end{figure}

Next, we take a closer look at the comparison of the model with the experiment shown in Fig.~\ref{fig:QPImethod}~(e). For $E-E^\mathrm{exp}_\mathrm{F}\approx\qtyrange{-0.5}{-0.1}{\eV}$ much less spectral weight along the $\overline{\Gamma K}$ compared to the $\overline{\Gamma M}$ direction is found, an effect not captured by the model. Neither is this reduction in spectral weight found in the first-principles calculations. To understand the difference between theory and experiment we explicitly simulate the experimental data as described by the joint density of states (JDOS), which is defined as
\begin{equation}
    \mathrm{JDOS}({\bf q}) = \int\mathrm{d}^2k\int_{E-\delta_E}^{E+\delta_E}\mathrm{d}E' \ I({\bf k}; E') I({\bf k}+{\bf q}; E'),
    \label{eq:JDOS}
\end{equation}
where $\delta_E$ defines an energy integration window for the intensities $I({\bf k}; E)$ of the constant energy contours at energies $E$. The intensities account for the wavefunction's weight in the vacuum where the STM tip scans the surface. In Fig.~\ref{fig:qpivsjdos}~(b) we plot the JDOS for ${E-E^\mathrm{exp}_\mathrm{F}}=\qty{-0.2}{\eV}$ together with the experimental pattern in Fig.~\ref{fig:qpivsjdos}~(a). As expected, the quasi-circular JDOS in panel (b) contrasts the experimental flower-like pattern in (a).
In Bi$_2$Se$_3$ similar differences between FTs of QPI and the JDOS are often attributed~\cite{Beidenkopf2011} to suppressed QPI in the $\overline{\Gamma K}$ direction due to forbidden back-scattering in the spin-momentum locked TSS.
At the surface of  PSBS, however, as discussed above, we have contributions from two bands with Rashba-like splitting where inter-band backscattering is allowed. Since the bandstructure of PSBS is not the origin of the observed flower-like scattering pattern, we recall that QPI is given by the convolution of the scattering potential with the initial and final density of states. When the wavelength of the scattered quasiparticles is large compared to the size of the potential, using the JDOS to approximate the FT of the QPI is justified. However, as we will see below, in the case of QPI on PSBS we need to include the scattering potential in our calculations. Specifically, in terms of Green's functions, the change in the charge density around a defect (scattering potential) is given as \cite{Ruessmann2020}
\begin{eqnarray}
    \Delta \rho({\bf r}; E) &=& -\frac{1}{\pi}\mathrm{Im}\mathrm{Tr} \int\mathrm{d}^3r'\int\mathrm{d}^3r''\ G^\mathrm{host}({\bf r}, {\bf r'}; E) \nonumber \\
        && \times  \Delta V({\bf r'}, {\bf r''}; E) G^\mathrm{imp}({\bf r''}, {\bf r}; E),
    \label{eq:Deltarhoimp}
\end{eqnarray}
where the impurity Green's function is connected to the host's Green's function via the Dyson equation,
\begin{equation}
    G^\mathrm{imp} = G^\mathrm{host} + G^\mathrm{host} \Delta V G^\mathrm{imp},
\end{equation}
and where the difference in the potential $\Delta V({\bf r})=V^\mathrm{imp}({\bf r})-V^\mathrm{host}({\bf r})$ around the impurity is used. While the exact form of the scattering potential is beyond the scope of this paper, it is instructive to approximate $\Delta \rho({\bf r})$ around subsurface $\mathrm{Pb}_\mathrm{Bi}$ defects in Bi$_2$Se$_3$ with their typical triangular shape observed at the surface in the STM (Fig.~\ref{fig:DefectSTM} and Appendix~\ref{app:scatteringpotential}).

Specifically, we use a simple shape function  [$T({\bf r})$] where the FT can be calculated analytically. Placing three scalar scatterers at the three vertices of an equilateral triangle with side length (2R) and one edge parallel to the $y$-axis, the FT reads~\cite{Stolyarov2021}
\begin{eqnarray}
\mathcal{F}_{T}(\mathbf{q},R) &=& \int_T \mathrm{d}^2r e^{i{\bf qr}} \nonumber \\
&\propto& e^{\frac{iRq_y}{\sqrt{3}}-iRq_x}+ e^{\frac{iRq_y}{\sqrt{3}}+iRq_x}+e^{\frac{-2iRq_y}{\sqrt{3}}}.\label{eq:F_3V}
\end{eqnarray}
We proceed to calculate the QDOS, which is defined as
\begin{equation}
    \mathrm{QDOS}(\mathbf{q}) =  \mathrm{JDOS}(\mathbf{q})|\mathcal{F}_T(\mathbf{q})|\approx\int\frac{\mathrm{d}^2r}{\sqrt{\Omega}} \Delta\rho(\mathbf{r}) e^{-i\mathbf{q}\cdot\mathbf{r}}
\end{equation} 
and show the comparison with the experiment in Fig.~\ref{fig:qpivsjdos} for $R=2b$, where $b$ is the in-plane lattice constant.
Clearly, the convolution with the scattering potential causes intensities along the $\overline{\Gamma K}$ ($\overline{\Gamma M}$) direction to be strongly suppressed (enhanced) resulting in the flower-like pattern similar to the experiment. Intuitively, the observation is understood as the modulation of the intensities by the Bragg conditions imposed by the triangular scatterer.

We note that the scattering potential generally decays faster than the measured QPI and excluding a region of the order of the Fermi wavelength $\lambda_F$ around each scatterer before calculating the FT is an established experimental technique~\cite{Chun2012} to suppress the influence of the potential. Unfortunately, this method cannot be applied in the case of PSBS due to the high defect density, resulting in an insufficient signal for analyzing the QPI pattern. Importantly, the relatively high defect density does not contradict the observed weak potential disorder (small $\Gamma$) in PSBS. The latter results from long-range fluctuations caused by bulk dopants—particularly $\mathrm{Bi}_\mathrm{Pb}$ antisites in the PbSe layers—leading to the surface puddles seen in Fig.~\ref{fig:fig1}~(g). In contrast, the QPI signal in Fig.~\ref{fig:QPImethod}~(a,c) reflects scattering from $\mathrm{Pb}_\mathrm{Bi}$ antisites in the Bi$_2$Se$_3$ surface layer (Appendix~\ref{app:scatteringpotential}).

\section{Summary}
STM topography of PSBS reveals a quasi-one-dimensional stripe pattern with a period of $\qty{2.1}{\nm}$, arising from a moiré effect between the hexagonal Bi$_2$Se$_3$ and square PbSe lattices, as confirmed by first-principles calculations. From STS, the bottom of the surface state is located at $E-E_\mathrm{F} \approx -0.65$~eV, and its spatial variations yield a potential disorder strength of $\Gamma = \qty{4}{meV}$—among the lowest reported for topological insulators.

QPI measurements over a broad energy range reveal a nearly parabolic dispersion of the dominant feature, consistent with prior ARPES~\cite{Nakayama2012,Nakayama2015,Nakayama2019} and theoretical results~\cite{Momida2018}. A simple model describes this with a Dirac velocity of $\qty{1e6}{\meter\per\second}$, weak warping ($\qty{26}{\eV\AA^3}$), inverted gap of $\qty{-1.24}{eV}$, and a symmetry-breaking potential of $\qty{0.05}{eV}$. The latter is visible in our first-principles calculations, manifesting as a Rashba-like splitting due to inversion symmetry breaking at the surface, though unresolved in the QPI.

The momentum distribution of the bias-dependent QPI shows a quasi-hexagonal pattern near $E_\mathrm{F}$ that evolves into a flower-like shape with suppressed intensity along $\overline{\Gamma K}$ at lower energies. While such anisotropy in 3D-TIs is often attributed to hexagonal warping, in PSBS it arises from the convolution of a nearly isotropic joint density of states with a threefold symmetric scattering potential, characteristic of atomic defects in Bi$_2$Se$_3$~\cite{Ruessmann2020}. A similar mechanism has been discussed for non-topological 2DEGs with triangular scatterers~\cite{mann2015commentary}.

\section*{Acknowledgements}

We are grateful to Raffaele Aliberti for his help with the calculated Friedel oscillations. PR thanks the Bavarian Ministry of Economic Affairs, Regional Development and Energy for financial support within High-Tech Agenda Project ``Bausteine f\"ur das Quantencomputing auf Basis topologischer Materialien mit experimentellen und theoretischen Ans\"atzen''. GB and PR gratefully acknowledges the computing time granted through JARA-HPC on the supercomputer JURECA at Forschungszentrum Jülich.
JB, MB, and YA  received funding from the European Research Council (ERC) under the European Union's Horizon 2020 research and innovation programme (grant agreement No 741121) and by the Deutsche Forschungsgemeinschaft (DFG, German Research Foundation) under CRC 1238 - 277146847 (Subprojects A04 and B06). MB, JB, PR, SB and YA are also grateful for funding by the DFG under Germany's Excellence Strategy - Cluster of Excellence Matter and Light for Quantum Computing (ML4Q) EXC 2004/1 - 390534769.

\section*{Data and Code Availability}

The dataset of our measurements and calculations is made publicly available in the materials cloud repository~\cite{dataset}. The source codes of the \texttt{Fleur} and \texttt{JuKKR} programs, and the AiiDA-KKR plugin~\cite{aiida-kkr-paper} are available as open-source repositories from Refs.~\onlinecite{Fleur, jukkr2022, aiida-kkr-code}.


\appendix
\section*{Appendix}
\section{Drift correction and symmetrization}
\label{app:Driftcor}
\begin{figure}[ht]
	\centering
	\includegraphics[width=\columnwidth]{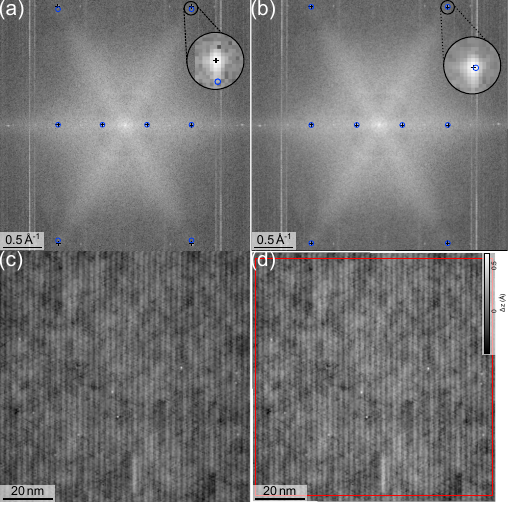}
	\caption{Drift correction. (a) FTs (a,b) of the topographies shown in (c,d). In (b,d) a linear transformation (details see text) was applied to compensate for thermal drift. (a,b) Blue circles and black crosses indicate the positions of various Bragg spots of the ideal lattice and from fitting a two-dimensional Gaussian to the experimental data, respectively. The red square in (d) indicates the area from which the QPI data is subsequently determined. Scale bars are \qty{0.5}{\text{\AA}^{-1}} (a,b) and  \qty{20}{\nm} (c,d).
	}
	\label{fig:Drift}
\end{figure}
It is well-known that thermal drift, piezoelectric creep, or calibration errors may distort the physical objects in the scanned image. For small (linear) distortions, a vector ${\bf r^\prime}=(x^\prime,y^\prime)$ in the undistorted space is related to the vector ${\bf r}=(x,y)$ in the (distorted) space via an affine transformation $M_r$ with $${\bf r^\prime}=M_r{\bf r}.$$
In the case of PSBS, determining the transformation $M_q$ in reciprocal space is most convenient. Specifically, we first determine Bragg spots $\left[{\bf q}_i=(q_{x_i},q_{y_i})\right]$ by fitting 2D Gaussians to the Fourier transform of the raw (distorted) experimental data. The results of the fits are superimposed as black crosses in Fig.~\ref{fig:Drift}~(a). Second, the ideal Bragg spots $\left[{\bf q^\prime}_i=(q_{x_i}^\prime,q_{y_i}^\prime)\right]$ are calculated based on the lattice constant determined from our first-principles calculations. The ${\bf q^\prime}_i$ are superimposed in Fig.~\ref{fig:Drift}~(a) as blue circles. Now, let the transformation be given as
$$M=\begin{pmatrix}
    M_{11} & M_{12} & M_{13}\\
    M_{21} & M_{22} & M_{23}\\
    0 & 0 & 1
\end{pmatrix},$$ then solving the matrix equation
\begin{equation}
    \begin{pmatrix}
        q_{x_1}^\prime\\
        q_{y_1}^\prime\\
        \vdots\\
        q_{x_i}^\prime\\
        q_{y_i}^\prime
    \end{pmatrix} =
    \begin{pmatrix}
        q_{x_1} & q_{y_1} & 1 & 0 & 0 & 0\\
        0 & 0 & 0 & q_{x_1} & q_{y_1} & 1\\
        \vdots & \vdots& \vdots& \vdots& \vdots & \vdots\\
        q_{x_i} & q_{y_i} & 1 & 0 & 0 & 0\\
        0 & 0 & 0 & q_{x_i} & q_{y_i} & 1
    \end{pmatrix}
    \begin{pmatrix}
        M_{11} \\
        M_{12} \\
        M_{13} \\
        M_{21} \\
        M_{22} \\
        M_{23} 
    \end{pmatrix},
\end{equation}
let us call it ${\bf P} = O \cdot {\bf M_q}$, in a least square sense gives us $M_q$ as ${\bf M_q}=O^{-1}{\bf P}$. Here $O^{-1}$ is the pseudoinverse of $O$. Applying $M_q$ to the raw data in Fig.~\ref{fig:Drift}~(a) leads to the corrected data shown in Fig.~\ref{fig:Drift}~(b). Correcting the real-space data is conveniently done with the real-space transformation $M_r=(M_q^T)^{-1}$. Examples of distorted and corrected real space data are shown in Fig.~\ref{fig:Drift}~(c,d). To minimize artifacts in the FTs of drift-corrected data the area outside the red square in Fig.~\ref{fig:Drift}~(d) is disregarded and only the periodic component~\cite{Moisan2011} of the FT is evaluated. Due to the high experimental stability in low-temperature STM measurements, distortions in our experimental data are generally minor. However, the transformations also conveniently align the FTs of different data taken with different tips on different sample surfaces with the orientation of the ideal lattice. Furthermore, we symmetrize our FT according to the 2mm symmetry of the surface-projected band structure to increase the signal-to-noise ratio. We note that there is no strict reason for the FT of the measured QPI data to have the same symmetry as the band structure. However, we have carefully checked for and excluded potential symmetrization artifacts from our analysis.

\section{Approximation of the scattering potential}
\label{app:scatteringpotential}

\begin{figure}[ht]
	\centering
	\includegraphics[width=0.9\linewidth]{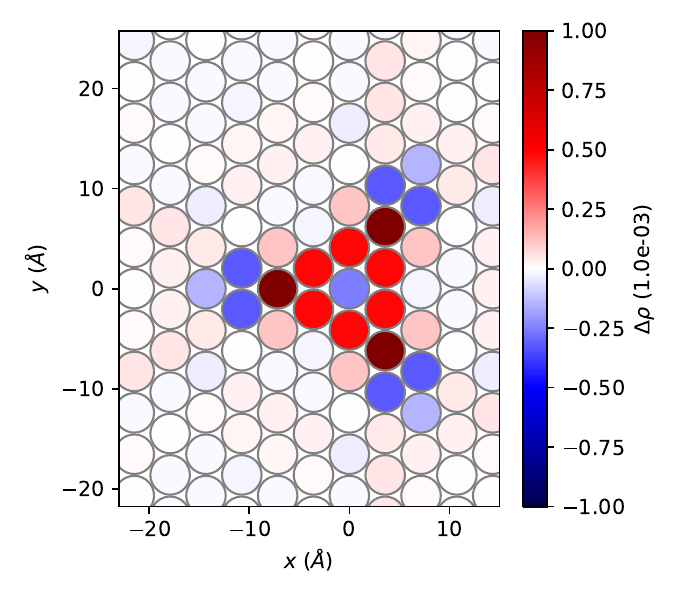}
	\caption{Friedel oscillations around a $\mathrm{Pb}_\mathrm{Bi}$-antisite defect in Bi$_2$Se$_3$. 
    Change in the charge density $\Delta\rho(\mathbf{r}, E^\mathrm{exp}_\mathrm{F})$ arising from a $\mathrm{Pb}_\mathrm{Bi}$ defect where Pb substitutes the second Bi atom from the surface in the Bi$_2$Se$_3$ quintuple layer structure. $\Delta\rho$ is calculated from Eq.~\eqref{eq:Deltarhoimp} in the vacuum at a distance of $z\approx3$\AA\ from the surface.
	}
	\label{fig:PbBidefect}
\end{figure}

Typical defects in the Bi$_2$Se$_3$ family of TI materials induce triangular Friedel oscillations in the charge density modulation around defects~\cite{Ruessmann2020}. To verify this for the $\mathrm{Pb}_\mathrm{Bi}$ that can occur in PSBS, we calculated the Friedel oscillations for Pb replacing the subsurface Bi atom in a three quintuple layer thick film of Bi$_2$Se$_3$. We use the \textit{ab initio} impurity embedding implemented in the relativistic full-potential JuKKR code \cite{jukkr2022} with an $\ell_\mathrm{max}=3$ cutoff in the angular momentum expansion and employ the local density approximation for the exchange correlation functional \cite{Vosko:80}.
The result is shown in Fig.~\ref{fig:PbBidefect} where the typical triangular shape of the order of a nanometer is visible. The subsurface location of the defect furthermore leads to a local minimum in the charge density at the center of the triangle.

\begin{figure}[ht]
	\centering
	\includegraphics[scale=0.9]{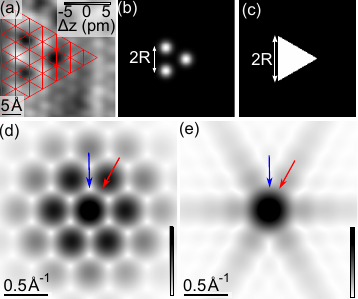}
	\caption{Scattering potential. (a) Zoom in around a $\mathrm{Pb}_\mathrm{Bi}$-antisite defect (Same data as in Fig.~\ref{fig:DefectSTM}, $I_0=\qty{2}{\nA}$, $V_0=\qty{600}{\mV}$). A standard high-pass filter was applied to enhance the contrast. The scale bar is \qty{0.5}{\nm}, and the Se-lattice is partially superimposed in red. Realspace schematic of the scattering potential for three scatters situated only at the vertices of an equilateral triangle (b) and for constant density throughout the triangle (c). (d,e) FT of (b,c), the scale bar is \qty{0.5}{\AA^{-1}}. The data in (c) is the same as in Fig.~\ref{fig:qpivsjdos}~(d). Red and blue arrows, respectively, highlight directions of high and low intensities.
	}
	\label{fig:ScatteringPot}
\end{figure}

Motivated by the characteristic triangular appearance of $\mathrm{Pb}_\mathrm{Bi}$-antisite defects in the experiment (Fig.~\ref{fig:DefectSTM}) and theoretical results of the Friedel oscillations around defects in $\mathrm{Bi}_2\mathrm{Se}_3$ (Fig.~\ref{fig:PbBidefect}) we used the approximation of three scattering sites at the vertices of an equilateral triangle described by Eq.~\ref{eq:F_3V}. Another shape approximating the experiment and for which an analytic expression of the FT is known is the triangle depicted in Fig.~\ref{fig:ScatteringPot}~(c). The FT for a constant density inside an equilateral triangle reads
\begin{eqnarray}
\mathcal{F}_T(\mathbf{q},R) &=& \int_T \mathrm{d}^2r e^{i{\bf qr}} \nonumber \\
&=& \frac{2\sqrt{3}e^{iq_xR\sqrt{3}}}{q_y(q_y^2-3q_x^2)}\big[q_ye^{iq_xR\sqrt{3}} \nonumber \\
&& -q_y\cos{(q_yR)}-i\sqrt{3}q_x\sin{(q_yR)}\big]
\end{eqnarray}
and is plotted in Fig.~\ref{fig:ScatteringPot}~(e) for $2R=7b$. Importantly, one sees again that intensity along $\overline{\Gamma M}$ ($\overline{\Gamma K}$) is high (low), illustrating that the influence of the scattering potential on the measured QPI is comparatively robust and independent of details of the shape function as long as the triangular symmetry and size $2R \sim \lambda_\mathrm{QPI}$ is captured.

\section{Band structure of the calculated film}
\label{app:states}

\begin{figure}[ht]
    \centering
    \includegraphics[width=1.\linewidth]{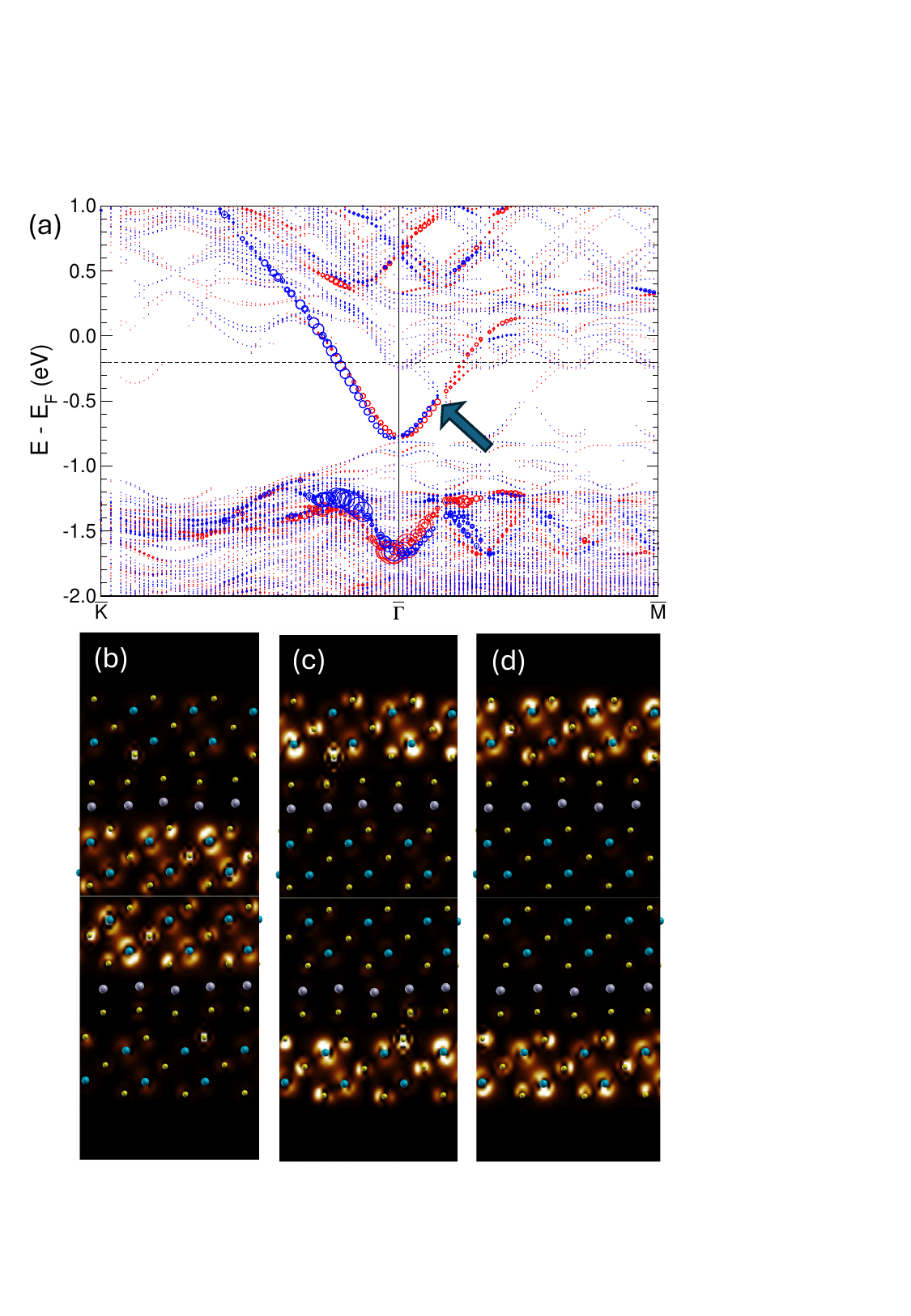}
    \caption{(a) Surface projected band structure of the calculated film unfolded to the hexagonal Bi$_2$Se$_3$ unit cell. The spin-polarization of the states is indicated by the size of the symbols, the direction (perpendicular to the $k$-vector) by the color. The experimental Fermi energy is indicated with the solid line for $E-E_\mathrm{F}$=\qty{-0.2}{eV}. The parabolic dispersion around $\overline{\Gamma}$, consists of several bands that are further analyzed at the position marked by the arrow: (b) Interface states of the inner Bi$_2$Se$_3$ bilayer with the neighboring PbSe layers (c) Interface states of the topmost Bi$_2$Se$_3$ layers with the PbSe layers below and (d) Surface states of the outer Bi$_2$Se$_3$ layers.}
    \label{fig:states}
\end{figure}

In Fig.~\ref{fig:states} the band structure of the Bi$_2$Se$_3$ / PbSe / 2(Bi$_2$Se$_3$) / PbSe / Bi$_2$Se$_3$ film is shown where the Brillouin zone was unfolded to the primitive hexagonal Bi$_2$Se$_3$ unit cell. The size of the symbols indicates the spin-polarization of the states in the topmost two atomic layers,  with respect to the axis perpendicular to the $k$-vector and the surface, and the color marks the direction. At $-1.7$~eV around $\overline{\Gamma}$ the Rashba-split surface state of Bi$_2$Se$_3$ can be seen; the parabolic feature starting at $-0.8$~eV consists of three doubly degenerate bands that are visualized for the position marked with the arrow in the panels below. They derive from the interface states already observed in the bulk calculations~\cite{Momida2018} that are localized at the inner (b) and outer (c) Bi$_2$Se$_3$/PbSe interface and the surface (d). These latter states are responsible for the quasiparticle interferences seen at the surface.

\section{Comparison of warping for PSBS, Bi$_2$Se$_3$ and Bi$_2$Te$_3$}

\begin{figure}[ht]
    \centering
    \includegraphics[width=1.\linewidth]{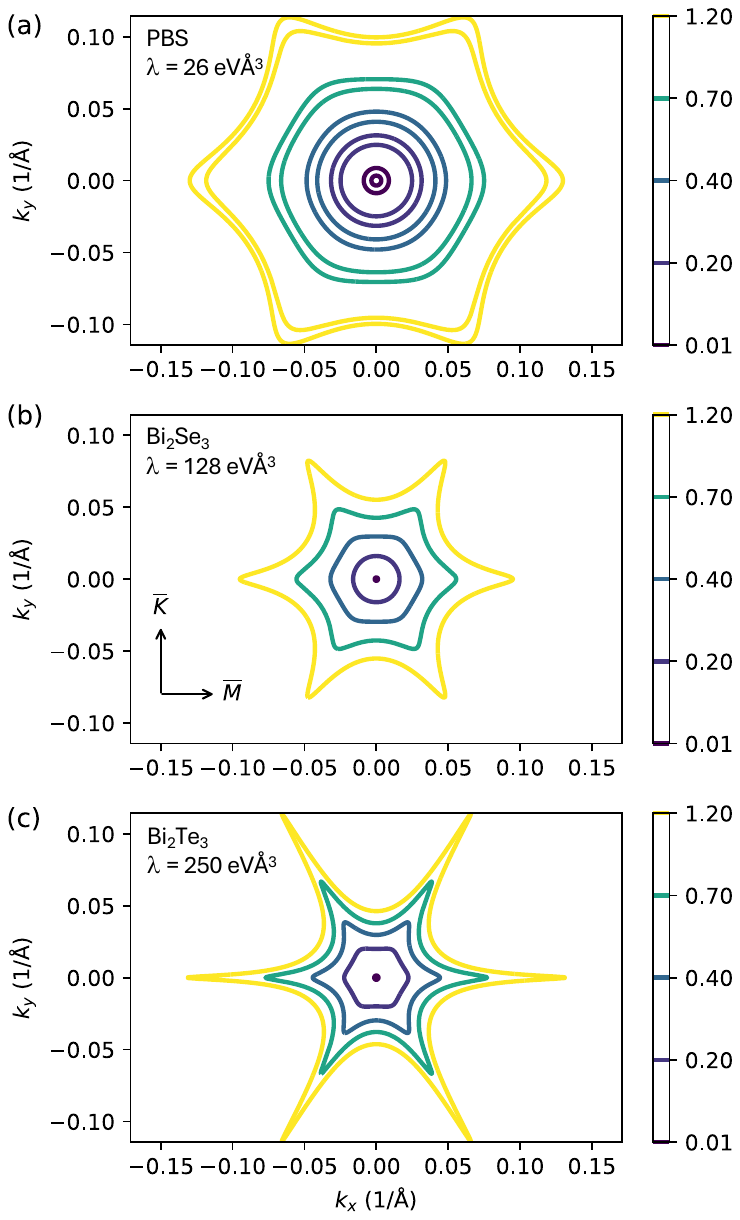}
    \caption{Constant energy contours for energies ranging from $0.01\,\mathrm{eV}$  to $1.2\,\mathrm{eV}$ showing the warping of the effective surface state model Hamiltonian \cite{Fu2009} included in Eq.~\eqref{eq:TSS} of the main text. The energy zero is the bottom of the TSS parabola (a) or the Dirac point (b,c). Compared are (a) data from PSBS of this work ($v_D=7.2\,\mathrm{eV\AA}$, $\lambda=26\,\mathrm{eV\AA^3}$, $\Delta=-1.24\,\mathrm{eV}$, $\tilde{V}_z = 0.05\,\mathrm{eV}$, $D=7.8\,\mathrm{eV\AA^2}$, $B=-8.4\,\mathrm{eV\AA^2}$), (b) data for clean Bi$_2$Se$_3$ ($v_D=3.55\,\mathrm{eV\AA}$, $\lambda=128\,\mathrm{eV\AA^3}$, $\tilde{V}=\Delta=0\,\mathrm{eV}$, $B=D=0\,\mathrm{eV\AA^2}$) \cite{Kuroda2010}, and (c) data for clean Bi$_2$Te$_3$ ($v_D=2.55\,\mathrm{eV\AA}$, $\lambda=250\,\mathrm{eV\AA^3}$, $\tilde{V}=\Delta=0\,\mathrm{eV}$, $B=D=0\,\mathrm{eV\AA^2}$) \cite{Fu2009}. The warping strength increases from (a-c).}
    \label{fig:warping}
\end{figure}

Figure~\ref{fig:warping} illustrates the strength of the warping term in PSBS, Bi$_2$Se$_3$ and Bi$_2$Te$_3$.

\section{Comparison of model Hamiltonian with experiment}
\label{app:ExperimentVSBHZmodel}

\begin{figure}[ht]
    \centering
    \includegraphics[width=.85\linewidth]{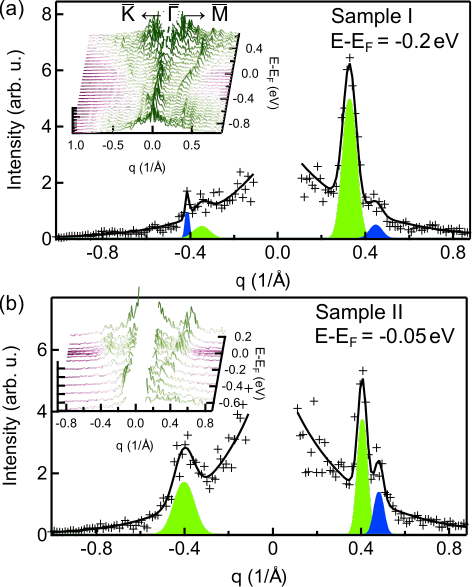}
    \caption{Momentum distribution curves along the $\overline{K} \leftarrow \overline{\Gamma} \rightarrow \overline{M}$ direction extracted from the symmetrized experimental E({\bf q}) data at $E-E_\mathrm{F}=\qty{-0.2}{eV}$ and -0.05~eV for sample I (a) and II (b), respectively. The inset in (a) shows the same data as Fig.~\ref{fig:QPImethod}~(e). (a,b) Data (crosses) fit (black line) with Gaussians and exponential decay, capturing dominant scattering vectors and smooth background intensity, respectively. The region $|q|<\qty{0.1}{\AA}$ has been excluded from the fit.}
    \label{fig:deconvolution}
\end{figure}

We measured the QPI for two surfaces (samples I and II). The symmetrized data for sample I are shown in Fig.~\ref{fig:QPImethod}, and the momentum distribution along the $\overline{\mathrm{K}\Gamma\mathrm{M}}$ direction is plotted in Fig.~\ref{fig:deconvolution}, together with the data for sample II. Complete sets of momentum distribution curves are shown in the insets, and we note an overall higher intensity of spectral features in the $\overline{\Gamma\mathrm{M}}$ direction compared to $\overline{\Gamma\mathrm{K}}$. Each curve was deconvoluted by combining Gaussian peaks and exponential decay, accounting for both dominant scattering vectors and a smooth background. The results of the deconvolution are explicitly shown for $E - E_\mathrm{F} = \qty{-0.2}{eV}$ (a, sample I) and \qty{-0.05}{eV} (b, sample II). Gaussians with high (green) and low (blue) intensity are attributed to scattering vectors of the surface state [labeled "1" in Fig.~\ref{fig:QPImethod}~(e)] and a spurious signal [labeled "2" in Fig.~\ref{fig:QPImethod}~(e)] due to the constant-current set-point effect~\cite{Macdonald_2016}, respectively. In short, the latter arises in constant-current mode, where the spatial variation of the tip height, $\Delta z({\bf r}) \sim \ln{\left(\int_{E_\mathrm{F}}^{eV_0} \mathrm{LDOS}({\bf r},E) , dE\right)}$, is convolved into the $\dd I({\bf r}, V_0)/\dd V$ measurement. This causes a contribution approximately given by the average of all $q$-values between $E_\mathrm{F}$ and $eV_0$ in the $\mathcal{F}_{\dd I/\dd V}({\bf q},V_0)$ data. However, we cannot exclude that some of the low-intensity features may also be related to interband scattering between the surface state and the bulk~\cite{Sunghun2011,Nagorkin2025}, or to disorder-related replicas~\cite{Beidenkopf2011}.

\begin{figure}[ht]
    \centering
    \includegraphics[width=.8\linewidth]{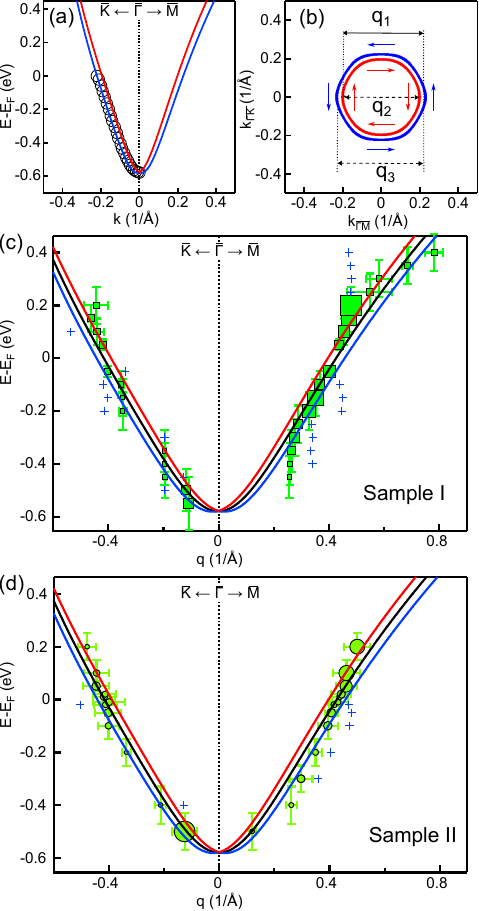}
    \caption{(a) Data (circles) extracted from Nakayama \etal~\cite{Nakayama2019} and the dispersion relation after Eq.~\ref{eq:TSS} (solid lines). (b) Fermi surface with interband ($q_1$) and intraband scattering vectors ($q_2$, $q_3$) indicated. Blue and red arrows indicate the spin direction. (c) Experimental scattering vectors (green squares and circles respectively for samples I and II, blue crosses are spurious features) extracted by deconvolution of our experimental data as shown in  Fig.~\ref{fig:deconvolution}. The size of the symbol indicates the weight (area) and the error bars indicate the HWHM. The error bars in E-E$_\mathrm{F}$ indicate the modulation voltage $V_\mathrm{mod}$.  Solid lines give the dispersion relations of $E(q_1)$ (black), $E(q_2)$ (red), and $E(q_3)$ (blue). $E_\mathrm{D}=\qty{-1.2}{eV}, v_\mathrm{D}=\qty{7.2}{eV\AA^{1}}, \tilde{V_z}=\qty{0.05}{eV}, \lambda=\qty{26}{eV\AA^{3}}, \Delta=\qty{-1.24}{eV}, D=\qty{7.8}{eV\AA^{2}}, B=\qty{-8.4}{eV\AA^{2}}$}
    \label{fig:ComparisionExpAndModel}
\end{figure}

Before comparing our experimental scattering vectors with the model (Eq.~\ref{eq:TSS}), we note that the dispersion relation is also compatible with the previous ARPES data by Nakayama \etal~\cite{Nakayama2019}. As shown in Fig.~\ref{fig:ComparisionExpAndModel}~(a), the data agree with the model parameters even though no Rashba-like splitting was reported. This is likely because the experimental momentum resolution ($\Delta k\approx\qty{0.01}{\AA^{-1}}$) is comparable to the splitting between the two helical branches.

Next, to compare the model with our QPI data, we first identify the three dominant scattering vectors $\textbf{q}_{1,2,3}$, as indicated in Fig.~\ref{fig:ComparisionExpAndModel}~(b). The corresponding dispersion relations $E(\textbf{q})$ follow the simple analytic form of Eq.~\ref{eq:TSS} with $k = q/2$ and $\tilde{V_z} = \qty{0}{eV}, \qty{-0.05}{eV}$, and \qty{0.05}{eV}, respectively. The vector $\textbf{q}_1$ corresponds to incoming and reflected waves with the same spin state, whereas $\textbf{q}_{2,3}$ involve opposite spin states. In the limit of single scattering, the orthogonality of spin states suppresses $\textbf{q}_{2,3}$, leaving only $\textbf{q}_1$ visible in the QPI~\cite{Ast2015}. Since $\textbf{q}_1$ is allowed in all directions, it cannot account for the observed anisotropy in scattering intensity between the $\overline{\Gamma\mathrm{M}}$ and $\overline{\Gamma\mathrm{K}}$ directions.

Considering multiple scattering, spin orthogonality may be relaxed, allowing $\textbf{q}_{2,3}$ to contribute in principle~\cite{Ast2015}. However, this scenario still does not result in directional anisotropy.

Finally, we account for the influence of the warping term, which induces out-of-plane spin polarization away from $\overline{\Gamma\mathrm{K}}$\cite{Fu2009}. This makes single scattering via $\textbf{q}_{2,3}$ possible along the $\overline{\Gamma\mathrm{M}}$ direction, while $\textbf{q}_1$ remains allowed in all directions. This mechanism, in principle, could introduce anisotropy in the QPI, consistent with experimental observations. However, as shown in Fig.~\ref{fig:warping}, the warping in PSBS is extremely weak and cannot explain the strong anisotropy observed at low energies. Specifically, at $E - E_\mathrm{F} = -0.2$eV, shown in Fig.~\ref{fig:qpivsjdos}, the warping is negligible. In summary, while the spin texture may enhance the intensity contrast between $\overline{\Gamma\mathrm{M}}$ and $\overline{\Gamma\mathrm{K}}$ at higher energies, it alone cannot explain the pronounced anisotropy observed down to $E - E_\mathrm{F} = -0.4$~eV.

\section{Interpolation scheme of computational constant energy contours}
\label{app:interpol}

\begin{figure*}[ht]
    \centering
    \includegraphics[width=0.8\linewidth]{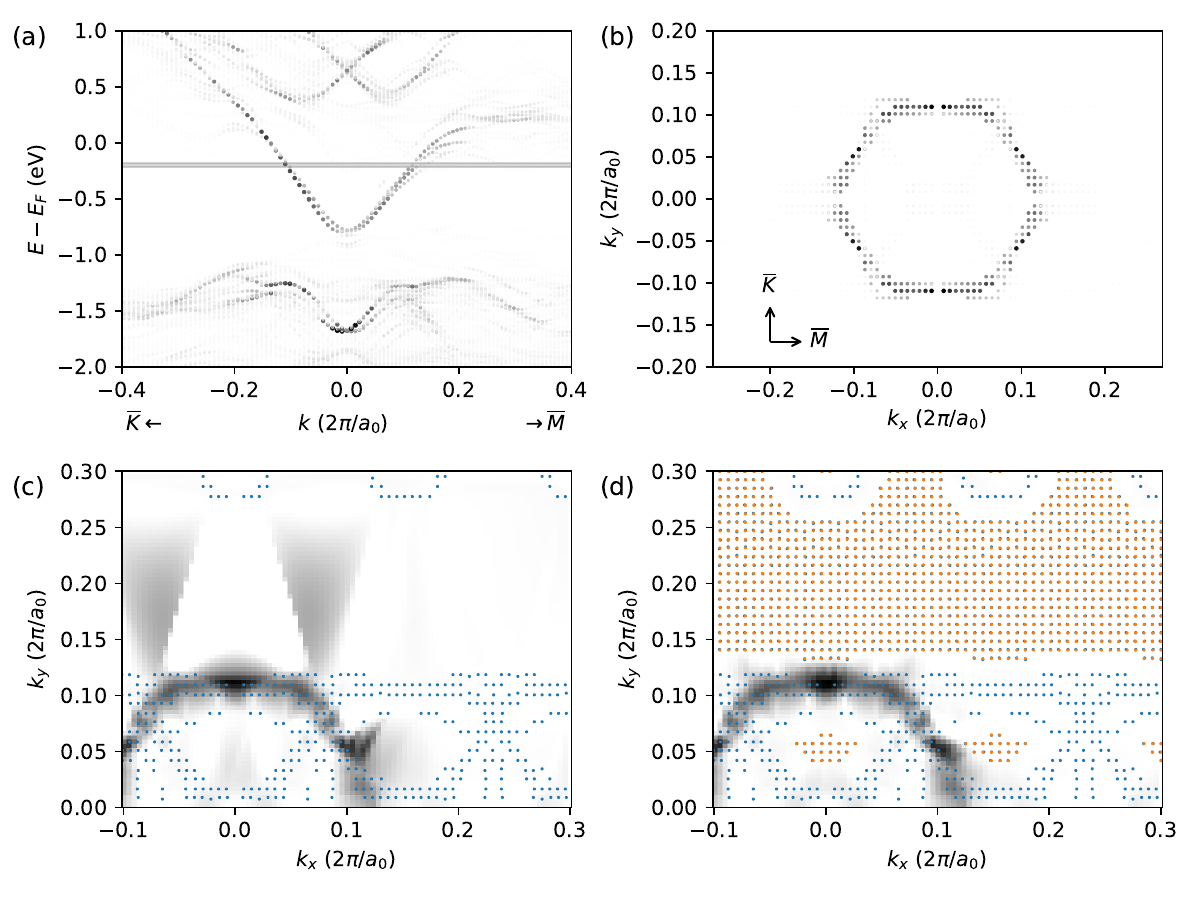}
    \caption{Constant energy contour extraction and interpolation scheme. (a) Band structure of PSBS around $\Gamma$ where the grey marked region highlights the energy window $|E-E_\mathrm{F}^\mathrm{theo}|\leq 0.02$eV that is integrated for the constant energy contour (CEC) in (b). Note that the bandstructure and corresponding CEC are unfolded to the larger Brillouin zone of Bi$_2$Se$_3$. Panels (c,d) demonstrate the interpolated CEC images using cubic splines, where the original data points are indicated in blue. In (d) additional data points (orange dots) with extra zeros are included to reduce spurious interpolation errors seen, for instance, as triangular shadows around $k_y\approx0.2\frac{2\pi}{a_0}$ in (c).  }
    \label{fig:interpolation}
\end{figure*}

The computed constant energy contours used in the JDOS analysis are computed on a $k$-point grid with 6080 points in the irreducible Brillouin zone. From this, we interpolate the constant energy contours (summarized in Fig.~\ref{fig:interpolation}) onto a finer grid used for plotting shown in Fig.~\ref{fig:qpivsjdos} of the main text.

\end{document}